\documentclass[pre,reprint]{revtex4-1}
\usepackage{graphicx,amsfonts,amsmath}
\newcommand{\be}{\begin{equation}}
\newcommand{\ee}{\end{equation}}
\usepackage{natbib}

\begin{document}

\title{Kuramoto variables as eigenvalues of unitary matrices} 



\author{Marcel Novaes$^{1,2}$ and Marcus A. M. de Aguiar$^2$}
\affiliation{$^1$Universidade Federal de Uberlândia, Uberlândia, MG, 38408-100, Brazil}
\affiliation{$^2$Universidade Estadual de Campinas, Campinas, SP, 13083-859, Brazil}

\date{\today}

\begin{abstract}
We generalize the Kuramoto model by interpreting the $N$ variables on the unit circle as eigenvalues of a $N$-dimensional unitary matrix $U$, in three versions: general unitary, symmetric unitary and special orthogonal. The time evolution is generated by $N^2$ coupled differential equations for the matrix elements of $U$, and synchronization happens when $U$ evolves into a multiple of the identity. The Ott-Antonsen ansatz is related to the Poisson kernels that are so useful in quantum transport, and we prove it in the case of identical natural frequencies. When the coupling constant is a matrix, we find some surprising new dynamical behaviors.  
\end{abstract}


\maketitle 

\section{Introduction}

Several natural and artificial systems rely on the synchronization of their units to function or to respond to external stimuli. Examples are pacemaker cells in the heart \cite{osaka2017}, neurons in brain \cite{cumin2007generalising,bhowmik2012well,ferrari2015phase,reis2021bursting}, and groups of real \cite{Buck1976} and artificial \cite{avila2003synchronization} fireflies. One of the most studied models of synchronization was proposed by Kuramoto \cite{Kuramoto1975}, and considers the joint evolution of $N$ coupled oscillators described only by their phases. 

The Kuramoto model has been extended and generalized in many ways over the years, including different types of coupling functions \cite{hong2011kuramoto,yeung1999time,breakspear2010generative}, networks of connections \cite{Rodrigues2016,Joyce2019},
special distributions of natural frequencies leading to explosive synchronization \cite{Gomez-Gardenes2011,Ji2013}, external driving forces
\cite{Childs2008,moreira2019global,moreira2019modular}, multidimensional systems \cite{2019continuous,Fariello2024a}, higher order interactions \cite{battiston2021physics,skardal2020higher} and complexified oscillators \cite{Seungjae2024}.  

Here we introduce a further generalization of the Kuramoto model which is defined in terms of a unitary matrix $U$. If $U$ is diagonal, each entry corresponds to a phase and the original Kuramoto model is recovered. However, for general matrices the dynamics becomes much richer, specially when the coupling $K$ is also extended to a matrix. 

Each oscillator in the Kuramoto model satisfies the first order differential equation, 
\be \dot{\theta}_j=\omega_j+\frac{K}{N}\sum_{k=1}^N\sin(\theta_k-\theta_j).\ee
Here $\{\omega_1,\ldots,\omega_N\}$ are the natural frequencies and $K$ measures the strength of the coupling. The initial conditions for the oscillators, $\{\theta_1(0),\ldots,\theta_N(0)\}$ are usually assumed to be uniformly distributed in the interval $(0, 2\pi)$, while the natural frequencies are drawn at random from some distribution $g(\omega)$.

Synchronization is usually measured by the so-called order parameter, defined as
\be\label{Z} Z=re^{i\psi}=\frac{1}{N}\sum_{k=1}^N e^{i\theta_k}.\ee
The modulus of $Z$ contains the essential information about the possibility of synchronization. If the phases are approximately uniformly distributed around the unit circle, then we expect that $|Z|\approx 1/\sqrt{N}$, whereas if the phases bunch up and become all approximately equal, $\theta_j\approx \psi$, then $|Z|\approx 1$.

In the limit of infinitely many oscillators, the order parameter can be written as the average value 
\be Z(t)=\int e^{i\theta}\rho(\theta,t)d\theta,\ee
where $\rho(\theta,t)$ is the distribution of phases at time $t$. This is like a ``circular moment'' of this distribution, with the higher circular moments, also known as
Kuramoto–Daido order parameters \cite{daido1996onset}, being its further Fourier coefficients,
\be Z_n(t)=\int e^{in\theta}\rho(\theta,t)d\theta.\ee
In a very influential paper, Ott and Antonsen suggested the ansatz \cite{Ott2008}
\be Z_n(t)=(Z(t))^n.\ee
This corresponds to assuming that the distribution $\rho(\theta,t)$ has the specific form
\be\label{form} \rho(\theta,t)=\frac{1}{2\pi}\frac{1-|Z|^2}{|1-\bar{Z}e^{i\theta}|^2}.\ee

Under the Ott-Antonsen ansatz, and assuming a Lorentzian distribution of the $\omega$'s, with half width $\Delta$, it can be shown that the order parameter satisfies the equation of motion
\be\label{param} \dot{Z}=-\Delta Z+\frac{K}{2}(1-|Z|^2)Z.\ee So the dynamics is reduced from a space of real dimension $N$ to a single complex number.  

In this paper we generalize the Kuramoto model for three classes of unitary matrices, namely general unitary matrices, symmetric unitary matrices and real unitary (orthogonal) matrices. The coupling constant is also interpreted as a coupling matrix, allowing for richer dynamics \cite{buzanello2022matrix}. Each of these classes has a reproducing kernel that is the Poisson kernel of the corresponding matrix space and is the equivalent of the Ott-Antonsen ansatz. We show numerical simulations of synchronization, where the coupling matrix is proportional to the identity, and more interesting cases using tri-diagonal coupling matrices. Finally we show that, for identical oscillators, the Ott-Antonsen ansatz does provide exact solutions for the order parameter. 

In Section \ref{2} we introduce a generalization of this model which is defined in terms of unitary matrices. In Section \ref{3}, we discuss how a modified Ott-Antonsen ansatz corresponds to the oscillators being distributed according to Poisson kernels. In Section \ref{4} we investigate numerically the onset of synchronization in the generalized model. In Section \ref{5} we prove that the ansatz is true when all natural frequencies are equal. 

\section{Unitary matrix evolution}\label{2}

If we define $x_j=e^{i\theta_j}$, then the order parameter can be used to write the dynamics in the form
\be \dot{x}_j=i\omega_j x_j+\frac{KZ}{2}-\frac{K\bar{Z}x_j^2}{2}.\ee Notice that the coupling between the different dynamical variables is induced by the presence of $Z$ in each equation of motion ($\bar{Z}$ is the complex conjugate).

We interpret the complex numbers $x_j$, which all have unit modulus, as the $N$ eigenvalues of a unitary matrix $U$. This suggests a generalization of the Kuramoto model in which the dynamics is not defined in terms of $\{x_1,\ldots,x_N\}$, but as a single equation of motion for $U$ (notice that we are not introducing a network of many interacting unitary matrices as in \cite{lohe2009non,ha2019emergent,bronski2020matrix}).

One natural generalization is
\be\label{model1} \dot{U}=i\Omega U+\frac{1}{2}KZ-\frac{1}{2}UK^\dagger U \bar{Z},\ee where $\Omega$ contains the natural frequencies, the order parameter is
\be Z=\frac{1}{N}{\rm Tr}(U),\ee and there are several possibilities for the coupling $K$, which is now a matrix: for example, it can be a multiple of the identity,  diagonal, real, symmetric, generic, etc. For any $K$, (\ref{model1}) leads to $\dot{U}U^\dagger=-U\dot{U}^\dagger$, so the unitarity constraint $UU^\dagger$ is satisfied for all times.  

Let $\mathcal{U}(N)$ be the group of complex $N$-dimensional unitary matrices. It comes equipped with a natural (Haar) probability distribution \cite{mehta2004random}, $P(U)$, which is invariant under left and right multiplication, $P(U)=P(UV)=P(VU)$. The initial condition $U(0)$ is drawn at random from this probability space, so the system has rotation invariance.

Without any loss of generality, we can assume that $\Omega$ is a diagonal matrix. If $K$ is a multiple of the identity, an initially diagonal $U$ remains diagonal for all times and this model reduces exactly to the Kuramoto model. However, the probability that a random initial $U(0)$ be diagonal is zero, as the space of diagonal matrices has no measure inside $\mathcal{U}(N)$. Therefore, for generic initial conditions we have a generalization of the Kuramoto model, consisting of $N^2$ coupled equations for the complex matrix elements,
\be \dot{U}_{jk}=i\Omega_jU_{jk}+\frac{1}{2}K_{jk}Z-\frac{1}{2}(UK^\dagger U)_{jk}\bar{Z}.\ee
Unitarity implies that the matrix elements are not independent. Instead, it is known that $\mathcal{U}(N)$ has real dimension $N^2$.

Synchronization happens in this model when $U(t)$ evolves in time to become not only diagonal, but a multiple of the identity, so that all eigenvalues approximately coincide, $U(\infty)\approx e^{i\psi}1_N$ and $|Z(\infty)|\approx 1$.

We could also consider $U$ to be symmetric. The space of unitary symmetric matrices also has a natural invariant probability measure and is known as the Circular Orthogonal Ensemble (COE) in the theory of random matrices \cite{mehta2004random}, where it is used to model quantum propagators and scattering matrices in the presence of time-reversal symmetry. In this case, a natural generalization is
\be\label{model2} \dot{U}=i\frac{\Omega U+U\Omega }{2}+\frac{1}{2}KZ-\frac{1}{2}U\bar{K} U\bar{Z},\ee with $K$ symmetric. It is easy to see that this equation preserves both unitarity and symmetry. Symmetry obviously reduces dimension: the space $COE(N)$ has real dimension $N(N+1)/2$.

And $U$ might also be taken real, therefore orthogonal. The space of special orthogonal matrices, $SO(N)$, has real dimension $N(N-1)/2$. When $N$ is odd, $+1$ is a fixed eigenvalue with no dynamics, so we restrict our attention to even $N$ in this case. Complex eigenvalues come in conjugate pairs, so the dynamics displays a $\mathbb{Z}_2$ symmetry about the real axis instead of the usual rotation symmetry.   A natural generalization of the Kuramoto model is
\be\label{model3} \dot{U}=\Omega U+\frac{1}{2}KZ-\frac{1}{2}UK^TUZ,\ee with $K$ and $Z$ real, and $\Omega$ block diagonal with $2\times 2$ blocks, i.e. 
\be \Omega=\bigoplus_j \begin{pmatrix} 0&\omega_j\\-\omega_j&0\end{pmatrix}.\ee
As a consequence of the $\mathbb{Z}_2$ symmetry, when the mean natural frequency is zero, complete phase locking can only happen around $\theta=0$ or $\theta=\pi$. Another possibility is the presence of two clusters of synchronized oscillators, one around some $\theta_0$ and the other around $-\theta_0$.

The eigenvalues of random unitary matrices are correlated and display repulsion. Their joint probability distribution \cite{meckes2019random} is proportional to 
\be\label{jpdf} \prod_{j<k}|x_j-x_k|^{\beta},\ee for $\mathcal{U}(N)$ and $COE(N)$, with  $\beta=2$ and $\beta=1$, respectively, and is proportional to 
\be\label{jpdfo} \prod_{j<k}|{\rm Re}(x_j)-{\rm Re}(x_k)|^{2}\ee
for $SO(2N)$. 

These are therefore the eigenvalue distributions of the initial matrix $U(0)$, depending on the model being used. In all cases, the density of initial phases is uniform in the interval $(0,2\pi)$.

\section{Ott-Antonsen ansatz and Poisson kernels}\label{3}

The initial values of the circular moments are all equal to zero when the variables $x_j$ are have the joint distributions discussed in the previous section, $Z_{n>0}(0)=0$. At time $t$, the Ott-Antonsen ansatz can be written as 
\be\label{ansatz} \frac{1}{N}{\rm Tr}(U^n)=\left( \frac{1}{N}{\rm Tr}(U)\right)^n=Z^n.\ee
Since the dynamics of our model is not identical to the Kuramoto model, it is not obvious that this ansatz is still valid, but we will check it numerically in the next section and prove it in a special case in Section V. 

If we introduce $\widetilde{Z}$ a multiple of the identity,
\be \widetilde{Z}=Z1_N,\ee
then we can write (\ref{ansatz}) as ${\rm Tr}(U^n)={\rm Tr}(\widetilde{Z}^n)$. Therefore, the ansatz amounts to saying that, even though $U(t)$ is not in general a multiple of the identity, it behaves like one as far as traces of powers are concerned.

This corresponds to a very specific joint probability distribution for the eigenvalues of $U$, known as the Poisson kernel. For $\mathcal{U}(N)$ and $COE(N)$, a very complete theory can be found in Hua \cite{hua1963harmonic}. For the special orthogonal group, it was derived much more recently in \cite{beri2009random}, because of a physical motivation: different Poisson kernels describe the distribution of the scattering matrix of quantum systems in different symmetry classes, when the contact points with the external world are not perfectly transparent (see \cite{mello1985information,doron1992some,baranger1996short,brouwer1995generalized,jarosz2015random}).

In all cases of interest to us, the kernel $\rho(\widetilde{Z},U)$ satisfies the reproducing property:
\be \int dU\rho(\widetilde{Z},U) {\rm Tr}(U^n)={\rm Tr}(\widetilde{Z}^n).\ee For the matrix ensembles we are considering, it is given by
\be \rho_{\mathcal{U}(N)}(\widetilde{Z},U)=\frac{\det(1_N-\widetilde{Z}\widetilde{Z}^\dagger)^{N}}{\det(1_N-U\widetilde{Z}^\dagger)^{2N}},\ee
\be \rho_{COE(N)}(\widetilde{Z},U)=\frac{\det(1_N-\widetilde{Z}\widetilde{Z}^\dagger)^{(N+1)/2}}{\det(1_N-U\widetilde{Z}^\dagger)^{N+1}},\ee
and
\be \rho_{SO(2N)}(\widetilde{Z},U)=\frac{\det(1_N-\widetilde{Z}\widetilde{Z}^T)^{2N-1}}{\det(1_N-U\widetilde{Z}^T)^{2N-1}}.\ee

\begin{figure*}[t]
\includegraphics[scale=0.6]{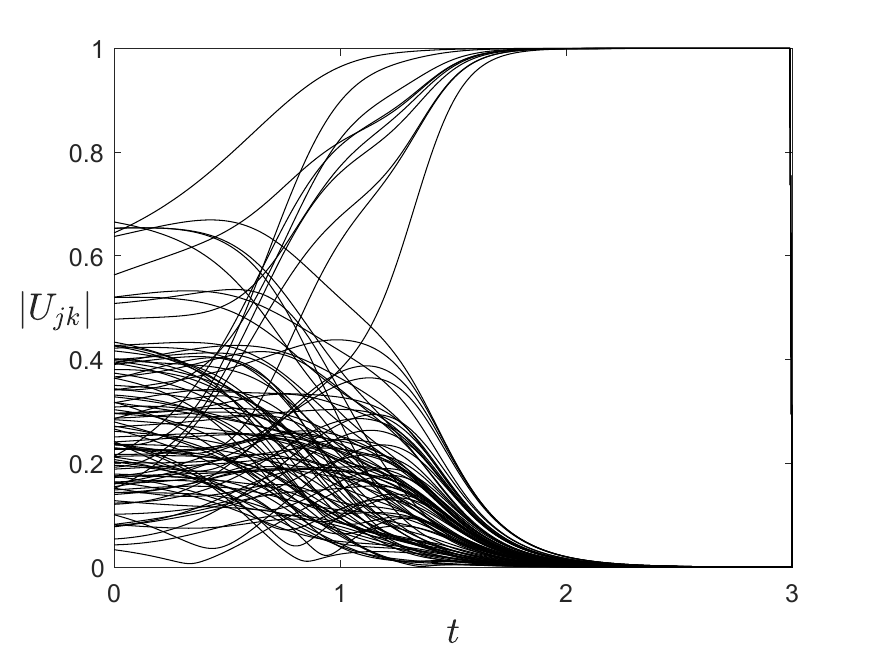}\includegraphics[scale=0.6]{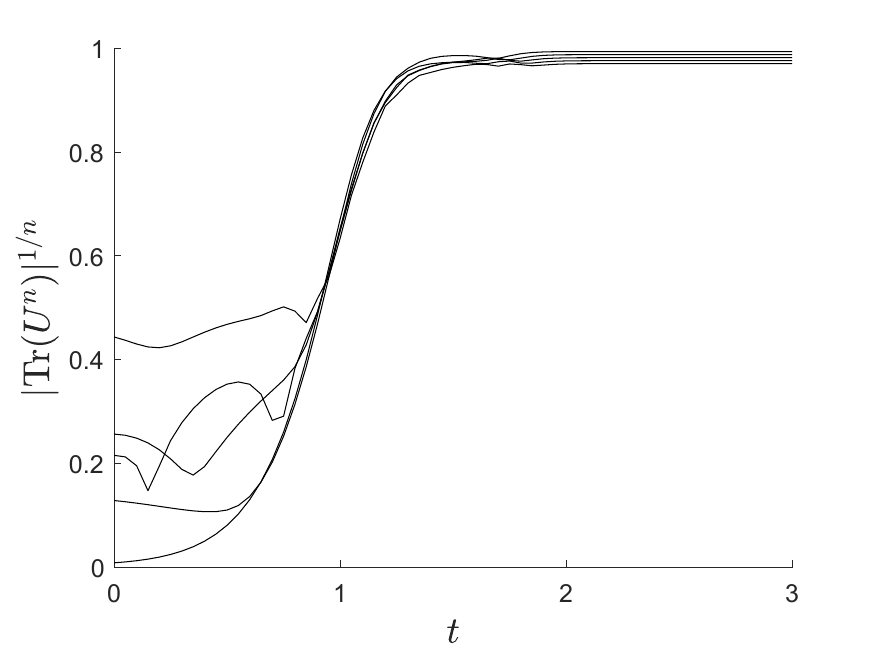}
\caption{Numerical simulation, as functions of time, with natural frequencies drawn at random from a normal distribution. a) Modulus of the matrix elements of $U$,  for $N=10$ and $K=5$. The off-diagonal ones tend to zero, and the diagonal ones tend to the unit circle. b) Modulus of the first few ($1\le n\le 5$) traces of powers of $U$,  for $N=100$ and $K=10$, rescaled by raising them to $1/n$. After an initial transient, they lie approximately on top of each other, corroborating the Ott-Antonsen ansatz.\label{Figure1}}
\end{figure*}

In the particular case that $\widetilde{Z}=Z1_N$ is a multiple of the identity, it can be written directly in terms of the eigenvalues (in the following expressions it must be remembered that $Z=\frac{1}{N}\sum_j x_j$). For  $\mathcal{U}(N)$, this is
\be\label{kernel1} \prod_{j=1}^N\frac{(1-|Z|^2)^N}{|1-\bar{Z}x_j|^{2N}}\prod_{j<k}|x_j-x_k|^{2},\ee
while for $COE(N)$ it is
\be\label{kernel2} \prod_{j=1}^N\frac{(1-|Z|^2)^{(N+1)/2}}{|1-\bar{Z}x_j|^{N+1}}\prod_{j<k}|x_j-x_k|,\ee
and for $SO(2N)$ it is
\be\label{kernelo} \prod_{j=1}^N\frac{(1-Z^2)^{2N-1}}{|1-Zx_j|^{2N-1}}\prod_{j<k}|{\rm Re}(x_j)-{\rm Re}(x_k)|^{2},\ee
all of which can be seen as generalizations of (\ref{form}).

Under the OA ansatz, as the matrix $U$ evolves under equations (\ref{model1}), (\ref{model2}) or (\ref{model3}), its eigenvalues are distributed, for all times, according to the corresponding Poisson kernel, namely  (\ref{kernel1}), (\ref{kernel2}) and (\ref{kernelo}), with time dependence coming from the parameter $Z(t)$.

In physics and in random matrix theory, Poisson kernels are used to model matrix ensembles, i.e. situations in which $U$ may be thought of as being a random variable. In our present approach to Kuramoto models, only $U(0)$ is chosen at random (from Haar measure, which is a Poisson kernel with $Z=0$), while the time evolution is deterministic. Nevertheless, when $N$ is large the eigenvalues of $U(t)$ can still the described by a distribution. 

Since the kernel in question is determined only by the symmetries imposed on $U$ and $Z(t)$, the system's dynamics is effectively reduced from a space of large dimension, comprising all matrix elements of $U$, to a single complex variable. As we have seen, the dynamics of the system takes place in real dimension $N^2$, $N(N+1)/2$ and $N(N-1)/2$ depending is $U$ is general unitary, symmetric or special orthogonal, so this dimensional reduction to a single complex number is even more dramatic than in the usual Kuramoto model.

\section{Numerical simulations}\label{4}

If the equations of motion (\ref{model1}), (\ref{model2}) and (\ref{model3}) could be integrated with infinite precision, the unitarity of $U$ would be guaranteed for all times. However, numerical integration requires small but finite time steps, and unitarity may degrade with time (in other words, the matrix spaces we consider are curved subspaces of the space of all matrices, and numerical integration does not exactly account for the curvature). We use a fourth order Runge-Kutta method and, at intermediate times, when we have $U^\dagger U=1+A$ with $A$ very small, we renormalize 
\be U\mapsto U(1+A)^{-1/2}\approx U(1-A/2),\ee
thereby imposing unitarity.

In Figure \ref{Figure1}, we show the results of numerical simulation, as functions of time, for a single random initial condition, with natural frequencies  drawn from a normal distribution with zero mean and unit variance. 

Panel a) shows the modulus of all matrix elements of $U$, for $N=10$ and the scalar coupling $K=5$.  The $90$ off-diagonal ones die down to zero, while the $10$ diagonal ones tend to the unit circle. So the time evolution effectively diagonalizes $U$. More than that, we have synchronization: all eigenvalues become approximately equal, as can be appreciated in panel b), because $|Z(t)|$ tends to $1$. We also plot the quantities $|{\rm Tr}(U^n)|^{1/n}$ for $1\le n\le 5$. The ansatz (\ref{ansatz}) being true implies that all these curves should lie on top of each other, and we see that this is a good approximation.

\begin{figure}[t]
\includegraphics[scale=0.6]{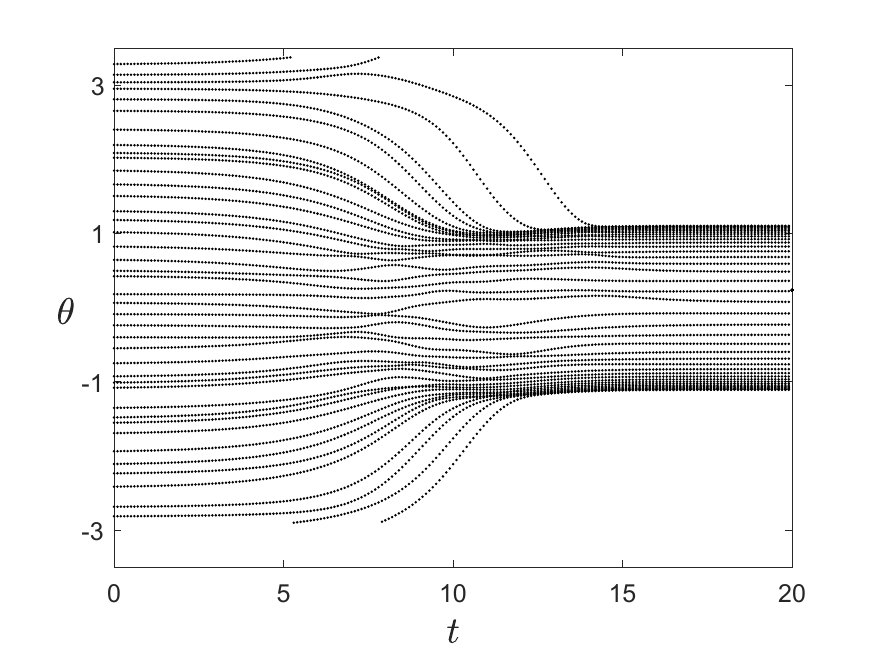}
\caption{Numerical simulation for $N=40$ with $\Omega=0$ and $K$ the nontrivial matrix in (\ref{Km}). The asymptotic eigenvalues of $U$ are determined by those of $K$, except for their midpoint. \label{Figure2}}
\end{figure}

\begin{figure}[t]
\includegraphics[scale=0.6]{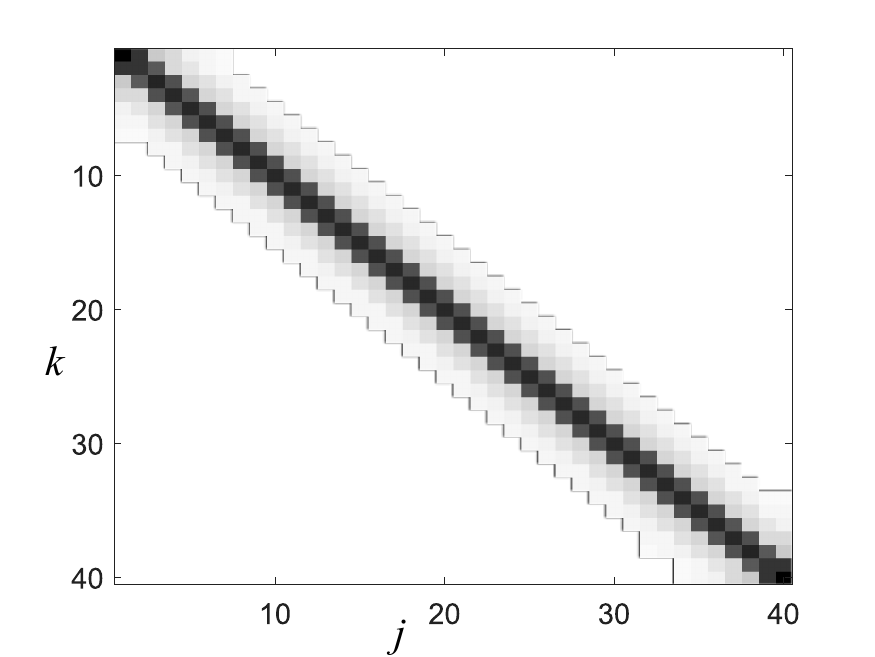}
\caption{Numerical simulation, with the same parameters as Figure \ref{Figure2}, of $|U_{jk}|(t)$, with white being zero. In this case $U$ does not diagonalize with time, but becomes banded.\label{Figure2b}}
\end{figure}

To show that the behavior of this model can be very different from the usual Kuramoto model, we consider vanishing natural frequencies, $\Omega=0$, but a  coupling $K$ which is a tridiagonal matrix, with diagonal and upper diagonal elements equal to 1, and lower diagonal elements equal to -1., i.e. 
\be\label{Km} K=\begin{pmatrix}
    1&1&0&\cdots\\
    -1&1&1&\\
    0&-1&1\\
    \vdots&&&\ddots
\end{pmatrix}\ee
We choose this coupling because the result is surprising. The eigenvalues of $U$ converge in time to a stationary state, which is regular but not synchronized. They are in fact given by
\be\label{xk} x_j=e^{i\phi}\frac{k_j}{|k_j|},\ee where $k_j$ are the eigenvalues of $K$ and $\phi$ depends on the initial condition $U(0)$. Since $K$ is tridiagonal and Toeplitz, its eigenvalues can be obtained exactly:
\be k_j=1-2i\cos\left(\frac{j}{N+1}\right).\ee
The spectral evolution of $U$ can be seen in Figure \ref{Figure2}, where we plot the results for $N=40$, with $\phi$ moved to zero for clarity. In this case $U$ does not diagonalize with time, but becomes banded, as we can see in Figure \ref{Figure2b}. A rigorous proof of this is a challenge. Notice that when $K$ is real, positive and diagonal, a relation like (\ref{xk}) would be equivalent to synchronization, so the behavior we observe for (\ref{Km}) can be seen as a generalization of the concept of synchronization.

\begin{figure}[t]
\includegraphics[scale=0.6]{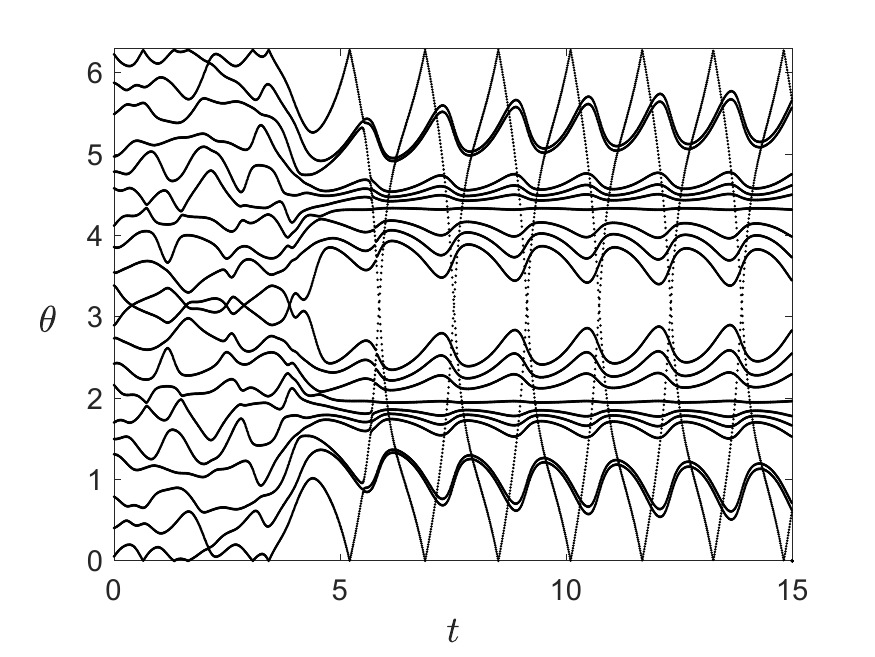}
\caption{Numerical simulation of the $SO(2N)$ model, for parameters given in the text. After some transient oscillations, the eigenvalues display intricate periodic behavior. Notice the $\mathbb{Z}_2$ symmetry of this model.\label{Figure3}}
\end{figure}

Another surprise was found when we simulated the model based on $SO(2N)$ with identical oscillators with $\omega_j=2$, and $K$ of the form $1_N\otimes \begin{pmatrix}
a&b\\-b&a \end{pmatrix}$. For example, we show in Figure \ref{Figure3} how the eigenvalues evolve, after an initial transient, into a nice and intricate periodic pattern, obtained with $a=6$ and $b=16$, for $N=20$. Again, a detailed understanding of such patterns is a challenge.

\section{Identical natural frequencies}\label{5}

In this Section we consider the special case when all natural frequencies are equal, and show that the OA ansatz is indeed true because of an underlying geometrical dynamics governed by linear fractional transformations, as has been demonstrated for the usual Kuramoto model \cite{pikovsky2008partially,marvel2009identical,chen2017hyperbolic}, except in the present case the variables are matrix valued. 

If all natural frequencies are equal, $\Omega=\omega 1_N$, we can write $U=U'e^{i\omega t}$ and the equation of motion becomes $\dot{U'}=\frac{1}{2}KZ'-\frac{1}{2}U'K^\dagger U' \bar{Z'}$, where $Z'=\frac{1}{N}{\rm Tr}U'$. So we might as well change to a rotating reference frame, or choose $\Omega=0$.

Let $B_N$ be the matrix unit ball defined by $1-XX^\dagger>0$. Its boundary is nothing but  $\mathcal{U}(N)$. The linear fractional transformation
\be M_{Y}(X)=(X+Y)(1+Y^\dagger X)^{-1}\ee maps $B_N$ into itself. Notice that 
$ M_Y(0)=Y.$ Expand for small $Yt$ to get 
\be M_{Yt}(X)\approx X+Yt-XY^\dagger Xt.\ee The derivative with respect to $t$ at $t=0$ is the generator of the transformation, and this is precisely in the form of our equation of motion,  (\ref{model1}).  

This means that if $X(t)\in B_N$ satisfies 
\be \dot{X}=i\omega X+\frac{1}{2}KZ-\frac{1}{2}XK^\dagger X \bar{Z},\ee then it must be of the form
\be X(t)=M_{Y(t)}(\xi(t)X_0),\ee for some time-dependent matrix parameters, a  rotation $\xi(t)\in\mathcal{U}(N)$ and a centroid $Y(t)\in B_N$. In particular, $Y(t)$ is the orbit of the origin and therefore satisfies itself the equation of motion,
\be \dot{Y}=\frac{1}{2}KZ-\frac{1}{2}YK^\dagger Y \bar{Z}.\ee Notice that $Y(0)$ is diagonal, being the zero matrix, and $Y(t)$ remains so for all times when $K$ is also diagonal. Moreover, if $K$ is a multiple of the identity, then so is $Y$.

The group generated by $(Y,\xi)$ acts transitively on $B_N$ and, instead of evolving points in time, we can evolve probability distributions. In other words, we can consider the push-forward of the initial Haar measure. Because of rotation invariance, the action of $\xi$ is irrelevant. On the other hand, the linear fractional transformations take the Haar measure precisely into the Poisson kernel \cite{hua1963harmonic}.

Therefore, the order parameter becomes the trace of the centroid, $Z(t)=\frac{1}{N}{\rm Tr}(Y(t))$, and the higher circular moments likewise satisfy $Z_n(t)=\frac{1}{N}{\rm Tr}((Y(t))^n)$.  We see that if $Y$ is a multiple of the identity, then the OA ansatz is indeed true. 

If $K$ is a multiple of the identity, complete dimensional reduction is accomplished and a single differential equation determines the evolution of the scalar order parameter: 
\be\label{final} \dot{Z}=\frac{KZ}{2}-\frac{\bar{K}\bar{Z}Z^2}{2}.\ee 
If $K$ is real,  we get back to (\ref{param}). If $K=K_r+iK_i$ has real and imaginary parts, the equations of motion for the modulus and phase of the order parameter become
\be \dot{r}=\frac{K_r r}{2}(1-r^2),\ee and
\be \dot{\psi}=\frac{K_i}{2}(1+r^2).\ee

In a slightly more generic case when $K$ is diagonal, the OA ansatz is no longer true for arbitray time (once synchronization is achieved, $U$ becomes a multiple of the identity and then the OA ansatz is trivially true), but we still have partial dimensional reduction, with $N$ (coupled) equations for the evolution of the $N$ complex elements of $Y$,
\be \dot{Y}_j=\frac{1}{2}K_jZ-\frac{1}{2}K_jY_j^2\bar{Z}.\ee We can still have synchronization in this case, depending on the values of the $K_j$. However, if $K$ is a more general real symmetric matrix there is no dimensional reduction, and typically no synchronization, as we have seen in the previous Section.

The calculations for $COE(N)$ and $SO(2N)$ are very similar. They are boundaries of  matrix balls invariant under linear fractional transforms and rotations, the generators of which coincide with our equations of motion (\ref{model2}) and (\ref{model3}). The pushforwards of the corresponding Haar measure are the Poisson kernels (\ref{kernel2}) and (\ref{kernelo}). When $K$ is a multiple of the identity, complete dimensional reduction takes place and the dynamics of the order parameter is again governed by (\ref{final}). 

If the natural frequencies are not all equal, the Poisson kernels can still be used as an approximation, but synchronizations no longer happens for any $K$. Instead, just like for the ordinary Kuramoto model, there is typically a critical value $K_c$ associated with a phase transition. Indeed, as the eigenvalue density corresponding to all three kernels is nothing but (\ref{form}), the original calculation by Ott and Antonsen can be repeated for a Lorentzian distribution of the $\omega$'s, with half width $\Delta$, and leads to $K_c=2\Delta$.

\section{Conclusions}

We have generalized the Kuramoto model to the evolution of the eigenvalues of a unitary matrix $U$, in such a way that the system is described by either a single matrix equation or by $N^2$ coupled scalar equations. By choosing $U$ to be symmetric, or real, different variants can be produced. We have related the Ott-Antonsen ansatz to matrix Poisson kernels, and showed that it is exact when the natural frequencies are identical.

This matrix version of the problem naturally admits a matrix valued coupling constant $K$. We have found numerically that specially for non-symmetric couplings this may lead to surprising dynamical behaviors, including for example an equilibrium state in which $U$ is banded with its spectrum being a renormalized version of the spectrum of $K$, or intricate periodic evolution for the eigenvalues, when $U$ is orthogonal.

These new dynamics suggest that matrix Kuramoto models may have very rich mathematical structures that are only being hinted at here and deserve further study. 



\begin{thebibliography}{41}%
\makeatletter
\providecommand \@ifxundefined [1]{%
 \@ifx{#1\undefined}
}%
\providecommand \@ifnum [1]{%
 \ifnum #1\expandafter \@firstoftwo
 \else \expandafter \@secondoftwo
 \fi
}%
\providecommand \@ifx [1]{%
 \ifx #1\expandafter \@firstoftwo
 \else \expandafter \@secondoftwo
 \fi
}%
\providecommand \natexlab [1]{#1}%
\providecommand \enquote  [1]{``#1''}%
\providecommand \bibnamefont  [1]{#1}%
\providecommand \bibfnamefont [1]{#1}%
\providecommand \citenamefont [1]{#1}%
\providecommand \href@noop [0]{\@secondoftwo}%
\providecommand \href [0]{\begingroup \@sanitize@url \@href}%
\providecommand \@href[1]{\@@startlink{#1}\@@href}%
\providecommand \@@href[1]{\endgroup#1\@@endlink}%
\providecommand \@sanitize@url [0]{\catcode `\\12\catcode `\$12\catcode
  `\&12\catcode `\#12\catcode `\^12\catcode `\_12\catcode `\%12\relax}%
\providecommand \@@startlink[1]{}%
\providecommand \@@endlink[0]{}%
\providecommand \url  [0]{\begingroup\@sanitize@url \@url }%
\providecommand \@url [1]{\endgroup\@href {#1}{\urlprefix }}%
\providecommand \urlprefix  [0]{URL }%
\providecommand \Eprint [0]{\href }%
\providecommand \doibase [0]{http://dx.doi.org/}%
\providecommand \selectlanguage [0]{\@gobble}%
\providecommand \bibinfo  [0]{\@secondoftwo}%
\providecommand \bibfield  [0]{\@secondoftwo}%
\providecommand \translation [1]{[#1]}%
\providecommand \BibitemOpen [0]{}%
\providecommand \bibitemStop [0]{}%
\providecommand \bibitemNoStop [0]{.\EOS\space}%
\providecommand \EOS [0]{\spacefactor3000\relax}%
\providecommand \BibitemShut  [1]{\csname bibitem#1\endcsname}%
\let\auto@bib@innerbib\@empty
\bibitem [{\citenamefont {Osaka}(2017)}]{osaka2017}%
  \BibitemOpen
  \bibfield  {author} {\bibinfo {author} {\bibfnamefont {M.}~\bibnamefont
  {Osaka}},\ }\href@noop {} {\bibfield  {journal} {\bibinfo  {journal} {Applied
  Mathematics}\ }\textbf {\bibinfo {volume} {8}},\ \bibinfo {pages} {1227}
  (\bibinfo {year} {2017})}\BibitemShut {NoStop}%
\bibitem [{\citenamefont {Cumin}\ and\ \citenamefont
  {Unsworth}(2007)}]{cumin2007generalising}%
  \BibitemOpen
  \bibfield  {author} {\bibinfo {author} {\bibfnamefont {D.}~\bibnamefont
  {Cumin}}\ and\ \bibinfo {author} {\bibfnamefont {C.}~\bibnamefont
  {Unsworth}},\ }\href@noop {} {\bibfield  {journal} {\bibinfo  {journal}
  {Physica D: Nonlinear Phenomena}\ }\textbf {\bibinfo {volume} {226}},\
  \bibinfo {pages} {181} (\bibinfo {year} {2007})}\BibitemShut {NoStop}%
\bibitem [{\citenamefont {Bhowmik}\ and\ \citenamefont
  {Shanahan}(2012)}]{bhowmik2012well}%
  \BibitemOpen
  \bibfield  {author} {\bibinfo {author} {\bibfnamefont {D.}~\bibnamefont
  {Bhowmik}}\ and\ \bibinfo {author} {\bibfnamefont {M.}~\bibnamefont
  {Shanahan}},\ }in\ \href@noop {} {\emph {\bibinfo {booktitle} {The 2012
  International Joint Conference on Neural Networks (IJCNN)}}}\ (\bibinfo
  {organization} {IEEE},\ \bibinfo {year} {2012})\ pp.\ \bibinfo {pages}
  {1--8}\BibitemShut {NoStop}%
\bibitem [{\citenamefont {Ferrari}\ \emph {et~al.}(2015)\citenamefont
  {Ferrari}, \citenamefont {Viana}, \citenamefont {Lopes},\ and\ \citenamefont
  {Stoop}}]{ferrari2015phase}%
  \BibitemOpen
  \bibfield  {author} {\bibinfo {author} {\bibfnamefont {F.~A.}\ \bibnamefont
  {Ferrari}}, \bibinfo {author} {\bibfnamefont {R.~L.}\ \bibnamefont {Viana}},
  \bibinfo {author} {\bibfnamefont {S.~R.}\ \bibnamefont {Lopes}}, \ and\
  \bibinfo {author} {\bibfnamefont {R.}~\bibnamefont {Stoop}},\ }\href@noop {}
  {\bibfield  {journal} {\bibinfo  {journal} {Neural Networks}\ }\textbf
  {\bibinfo {volume} {66}},\ \bibinfo {pages} {107} (\bibinfo {year}
  {2015})}\BibitemShut {NoStop}%
\bibitem [{\citenamefont {Reis}\ \emph {et~al.}(2021)\citenamefont {Reis},
  \citenamefont {Iarosz}, \citenamefont {Ferrari}, \citenamefont {Caldas},
  \citenamefont {Batista},\ and\ \citenamefont {Viana}}]{reis2021bursting}%
  \BibitemOpen
  \bibfield  {author} {\bibinfo {author} {\bibfnamefont {A.~S.}\ \bibnamefont
  {Reis}}, \bibinfo {author} {\bibfnamefont {K.~C.}\ \bibnamefont {Iarosz}},
  \bibinfo {author} {\bibfnamefont {F.~A.}\ \bibnamefont {Ferrari}}, \bibinfo
  {author} {\bibfnamefont {I.~L.}\ \bibnamefont {Caldas}}, \bibinfo {author}
  {\bibfnamefont {A.~M.}\ \bibnamefont {Batista}}, \ and\ \bibinfo {author}
  {\bibfnamefont {R.~L.}\ \bibnamefont {Viana}},\ }\href@noop {} {\bibfield
  {journal} {\bibinfo  {journal} {Chaos, Solitons \& Fractals}\ }\textbf
  {\bibinfo {volume} {142}},\ \bibinfo {pages} {110395} (\bibinfo {year}
  {2021})}\BibitemShut {NoStop}%
\bibitem [{\citenamefont {{John}}\ and\ \citenamefont
  {{Buck}}(1976)}]{Buck1976}%
  \BibitemOpen
  \bibfield  {author} {\bibinfo {author} {\bibnamefont {{John}}}\ and\ \bibinfo
  {author} {\bibfnamefont {E.}~\bibnamefont {{Buck}}},\ }\href {\doibase
  10.1038/scientificamerican0576-74} {\bibfield  {journal} {\bibinfo  {journal}
  {Scientific American}\ }\textbf {\bibinfo {volume} {234}},\ \bibinfo {pages}
  {74} (\bibinfo {year} {1976})}\BibitemShut {NoStop}%
\bibitem [{\citenamefont {{\'A}vila}\ \emph {et~al.}(2003)\citenamefont
  {{\'A}vila}, \citenamefont {Guisset},\ and\ \citenamefont
  {Deneubourg}}]{avila2003synchronization}%
  \BibitemOpen
  \bibfield  {author} {\bibinfo {author} {\bibfnamefont {G.~R.}\ \bibnamefont
  {{\'A}vila}}, \bibinfo {author} {\bibfnamefont {J.-L.}\ \bibnamefont
  {Guisset}}, \ and\ \bibinfo {author} {\bibfnamefont {J.-L.}\ \bibnamefont
  {Deneubourg}},\ }\href@noop {} {\bibfield  {journal} {\bibinfo  {journal}
  {Physica D: Nonlinear Phenomena}\ }\textbf {\bibinfo {volume} {182}},\
  \bibinfo {pages} {254} (\bibinfo {year} {2003})}\BibitemShut {NoStop}%
\bibitem [{\citenamefont {Kuramoto}(1975)}]{Kuramoto1975}%
  \BibitemOpen
  \bibfield  {author} {\bibinfo {author} {\bibfnamefont {Y.}~\bibnamefont
  {Kuramoto}},\ }in\ \href {\doibase 10.1007/BFb0013365} {\emph {\bibinfo
  {booktitle} {International Symposium on Mathematical Problems in Theoretical
  Physics}}}\ (\bibinfo  {publisher} {Springer-Verlag},\ \bibinfo {address}
  {Berlin/Heidelberg},\ \bibinfo {year} {1975})\ pp.\ \bibinfo {pages}
  {420--422}\BibitemShut {NoStop}%
\bibitem [{\citenamefont {Hong}\ and\ \citenamefont
  {Strogatz}(2011)}]{hong2011kuramoto}%
  \BibitemOpen
  \bibfield  {author} {\bibinfo {author} {\bibfnamefont {H.}~\bibnamefont
  {Hong}}\ and\ \bibinfo {author} {\bibfnamefont {S.~H.}\ \bibnamefont
  {Strogatz}},\ }\href@noop {} {\bibfield  {journal} {\bibinfo  {journal}
  {Physical Review Letters}\ }\textbf {\bibinfo {volume} {106}},\ \bibinfo
  {pages} {054102} (\bibinfo {year} {2011})}\BibitemShut {NoStop}%
\bibitem [{\citenamefont {Yeung}\ and\ \citenamefont
  {Strogatz}(1999)}]{yeung1999time}%
  \BibitemOpen
  \bibfield  {author} {\bibinfo {author} {\bibfnamefont {M.~S.}\ \bibnamefont
  {Yeung}}\ and\ \bibinfo {author} {\bibfnamefont {S.~H.}\ \bibnamefont
  {Strogatz}},\ }\href@noop {} {\bibfield  {journal} {\bibinfo  {journal}
  {Physical Review Letters}\ }\textbf {\bibinfo {volume} {82}},\ \bibinfo
  {pages} {648} (\bibinfo {year} {1999})}\BibitemShut {NoStop}%
\bibitem [{\citenamefont {Breakspear}\ \emph {et~al.}(2010)\citenamefont
  {Breakspear}, \citenamefont {Heitmann},\ and\ \citenamefont
  {Daffertshofer}}]{breakspear2010generative}%
  \BibitemOpen
  \bibfield  {author} {\bibinfo {author} {\bibfnamefont {M.}~\bibnamefont
  {Breakspear}}, \bibinfo {author} {\bibfnamefont {S.}~\bibnamefont
  {Heitmann}}, \ and\ \bibinfo {author} {\bibfnamefont {A.}~\bibnamefont
  {Daffertshofer}},\ }\href@noop {} {\bibfield  {journal} {\bibinfo  {journal}
  {Frontiers in human neuroscience}\ }\textbf {\bibinfo {volume} {4}},\
  \bibinfo {pages} {190} (\bibinfo {year} {2010})}\BibitemShut {NoStop}%
\bibitem [{\citenamefont {Rodrigues}\ \emph {et~al.}(2016)\citenamefont
  {Rodrigues}, \citenamefont {Peron}, \citenamefont {Ji},\ and\ \citenamefont
  {Kurths}}]{Rodrigues2016}%
  \BibitemOpen
  \bibfield  {author} {\bibinfo {author} {\bibfnamefont {F.~A.}\ \bibnamefont
  {Rodrigues}}, \bibinfo {author} {\bibfnamefont {T.~K. D.~M.}\ \bibnamefont
  {Peron}}, \bibinfo {author} {\bibfnamefont {P.}~\bibnamefont {Ji}}, \ and\
  \bibinfo {author} {\bibfnamefont {J.}~\bibnamefont {Kurths}},\ }\href
  {\doibase 10.1016/j.physrep.2015.10.008} {\bibfield  {journal} {\bibinfo
  {journal} {Physics Reports}\ }\textbf {\bibinfo {volume} {610}},\ \bibinfo
  {pages} {1} (\bibinfo {year} {2016})},\ \Eprint
  {http://arxiv.org/abs/1511.07139} {arXiv:1511.07139} \BibitemShut {NoStop}%
\bibitem [{\citenamefont {Climaco}\ and\ \citenamefont
  {Saa}(2019)}]{Joyce2019}%
  \BibitemOpen
  \bibfield  {author} {\bibinfo {author} {\bibfnamefont {J.~S.}\ \bibnamefont
  {Climaco}}\ and\ \bibinfo {author} {\bibfnamefont {A.}~\bibnamefont {Saa}},\
  }\href {\doibase 10.1063/1.5097847} {\bibfield  {journal} {\bibinfo
  {journal} {Chaos: An Interdisciplinary Journal of Nonlinear Science}\
  }\textbf {\bibinfo {volume} {29}},\ \bibinfo {pages} {073115} (\bibinfo
  {year} {2019})},\ \Eprint
  {http://arxiv.org/abs/https://doi.org/10.1063/1.5097847}
  {https://doi.org/10.1063/1.5097847} \BibitemShut {NoStop}%
\bibitem [{\citenamefont {Gomez-Gardenes}\ \emph {et~al.}(2011)\citenamefont
  {Gomez-Gardenes}, \citenamefont {Gomez}, \citenamefont {Arenas},\ and\
  \citenamefont {Moreno}}]{Gomez-Gardenes2011}%
  \BibitemOpen
  \bibfield  {author} {\bibinfo {author} {\bibfnamefont {J.}~\bibnamefont
  {Gomez-Gardenes}}, \bibinfo {author} {\bibfnamefont {S.}~\bibnamefont
  {Gomez}}, \bibinfo {author} {\bibfnamefont {A.}~\bibnamefont {Arenas}}, \
  and\ \bibinfo {author} {\bibfnamefont {Y.}~\bibnamefont {Moreno}},\ }\href
  {\doibase 10.1103/PhysRevLett.106.128701} {\bibfield  {journal} {\bibinfo
  {journal} {Physical Review Letters}\ }\textbf {\bibinfo {volume} {106}},\
  \bibinfo {pages} {1} (\bibinfo {year} {2011})},\ \Eprint
  {http://arxiv.org/abs/1102.4823} {arXiv:1102.4823} \BibitemShut {NoStop}%
\bibitem [{\citenamefont {Ji}\ \emph {et~al.}(2013)\citenamefont {Ji},
  \citenamefont {Peron}, \citenamefont {Menck}, \citenamefont {Rodrigues},\
  and\ \citenamefont {Kurths}}]{Ji2013}%
  \BibitemOpen
  \bibfield  {author} {\bibinfo {author} {\bibfnamefont {P.}~\bibnamefont
  {Ji}}, \bibinfo {author} {\bibfnamefont {T.~K.~D.}\ \bibnamefont {Peron}},
  \bibinfo {author} {\bibfnamefont {P.~J.}\ \bibnamefont {Menck}}, \bibinfo
  {author} {\bibfnamefont {F.~A.}\ \bibnamefont {Rodrigues}}, \ and\ \bibinfo
  {author} {\bibfnamefont {J.}~\bibnamefont {Kurths}},\ }\href {\doibase
  10.1103/PhysRevLett.110.218701} {\bibfield  {journal} {\bibinfo  {journal}
  {Physical Review Letters}\ }\textbf {\bibinfo {volume} {110}},\ \bibinfo
  {pages} {1} (\bibinfo {year} {2013})},\ \Eprint
  {http://arxiv.org/abs/arXiv:1303.3498v2} {arXiv:arXiv:1303.3498v2}
  \BibitemShut {NoStop}%
\bibitem [{\citenamefont {Childs}\ and\ \citenamefont
  {Strogatz}(2008)}]{Childs2008}%
  \BibitemOpen
  \bibfield  {author} {\bibinfo {author} {\bibfnamefont {L.~M.}\ \bibnamefont
  {Childs}}\ and\ \bibinfo {author} {\bibfnamefont {S.~H.}\ \bibnamefont
  {Strogatz}},\ }\href {\doibase 10.1063/1.3049136} {\bibfield  {journal}
  {\bibinfo  {journal} {Chaos}\ }\textbf {\bibinfo {volume} {18}},\ \bibinfo
  {pages} {1} (\bibinfo {year} {2008})},\ \Eprint
  {http://arxiv.org/abs/0807.4717} {arXiv:0807.4717} \BibitemShut {NoStop}%
\bibitem [{\citenamefont {Moreira}\ and\ \citenamefont
  {de~Aguiar}(2019{\natexlab{a}})}]{moreira2019global}%
  \BibitemOpen
  \bibfield  {author} {\bibinfo {author} {\bibfnamefont {C.~A.}\ \bibnamefont
  {Moreira}}\ and\ \bibinfo {author} {\bibfnamefont {M.~A.}\ \bibnamefont
  {de~Aguiar}},\ }\href@noop {} {\bibfield  {journal} {\bibinfo  {journal}
  {Physica A: Statistical Mechanics and its Applications}\ }\textbf {\bibinfo
  {volume} {514}},\ \bibinfo {pages} {487} (\bibinfo {year}
  {2019}{\natexlab{a}})}\BibitemShut {NoStop}%
\bibitem [{\citenamefont {Moreira}\ and\ \citenamefont
  {de~Aguiar}(2019{\natexlab{b}})}]{moreira2019modular}%
  \BibitemOpen
  \bibfield  {author} {\bibinfo {author} {\bibfnamefont {C.~A.}\ \bibnamefont
  {Moreira}}\ and\ \bibinfo {author} {\bibfnamefont {M.~A.}\ \bibnamefont
  {de~Aguiar}},\ }\href@noop {} {\bibfield  {journal} {\bibinfo  {journal}
  {Physica A: Statistical Mechanics and its Applications}\ }\textbf {\bibinfo
  {volume} {533}},\ \bibinfo {pages} {122051} (\bibinfo {year}
  {2019}{\natexlab{b}})}\BibitemShut {NoStop}%
\bibitem [{\citenamefont {Chandra}\ \emph {et~al.}(2019)\citenamefont
  {Chandra}, \citenamefont {Girvan},\ and\ \citenamefont
  {Ott}}]{2019continuous}%
  \BibitemOpen
  \bibfield  {author} {\bibinfo {author} {\bibfnamefont {S.}~\bibnamefont
  {Chandra}}, \bibinfo {author} {\bibfnamefont {M.}~\bibnamefont {Girvan}}, \
  and\ \bibinfo {author} {\bibfnamefont {E.}~\bibnamefont {Ott}},\ }\href@noop
  {} {\bibfield  {journal} {\bibinfo  {journal} {Physical Review X}\ }\textbf
  {\bibinfo {volume} {9}},\ \bibinfo {pages} {011002} (\bibinfo {year}
  {2019})}\BibitemShut {NoStop}%
\bibitem [{\citenamefont {Fariello}\ and\ \citenamefont
  {de~Aguiar}(2024)}]{Fariello2024a}%
  \BibitemOpen
  \bibfield  {author} {\bibinfo {author} {\bibfnamefont {R.}~\bibnamefont
  {Fariello}}\ and\ \bibinfo {author} {\bibfnamefont {M.~A.}\ \bibnamefont
  {de~Aguiar}},\ }\href@noop {} {\bibfield  {journal} {\bibinfo  {journal}
  {Chaos, Solitons \& Fractals}\ }\textbf {\bibinfo {volume} {179}},\ \bibinfo
  {pages} {114431} (\bibinfo {year} {2024})}\BibitemShut {NoStop}%
\bibitem [{\citenamefont {Battiston}\ \emph {et~al.}(2021)\citenamefont
  {Battiston}, \citenamefont {Amico}, \citenamefont {Barrat}, \citenamefont
  {Bianconi}, \citenamefont {Ferraz~de Arruda}, \citenamefont {Franceschiello},
  \citenamefont {Iacopini}, \citenamefont {K{\'e}fi}, \citenamefont {Latora},
  \citenamefont {Moreno} \emph {et~al.}}]{battiston2021physics}%
  \BibitemOpen
  \bibfield  {author} {\bibinfo {author} {\bibfnamefont {F.}~\bibnamefont
  {Battiston}}, \bibinfo {author} {\bibfnamefont {E.}~\bibnamefont {Amico}},
  \bibinfo {author} {\bibfnamefont {A.}~\bibnamefont {Barrat}}, \bibinfo
  {author} {\bibfnamefont {G.}~\bibnamefont {Bianconi}}, \bibinfo {author}
  {\bibfnamefont {G.}~\bibnamefont {Ferraz~de Arruda}}, \bibinfo {author}
  {\bibfnamefont {B.}~\bibnamefont {Franceschiello}}, \bibinfo {author}
  {\bibfnamefont {I.}~\bibnamefont {Iacopini}}, \bibinfo {author}
  {\bibfnamefont {S.}~\bibnamefont {K{\'e}fi}}, \bibinfo {author}
  {\bibfnamefont {V.}~\bibnamefont {Latora}}, \bibinfo {author} {\bibfnamefont
  {Y.}~\bibnamefont {Moreno}},  \emph {et~al.},\ }\href@noop {} {\bibfield
  {journal} {\bibinfo  {journal} {Nature Physics}\ }\textbf {\bibinfo {volume}
  {17}},\ \bibinfo {pages} {1093} (\bibinfo {year} {2021})}\BibitemShut
  {NoStop}%
\bibitem [{\citenamefont {Skardal}\ and\ \citenamefont
  {Arenas}(2020)}]{skardal2020higher}%
  \BibitemOpen
  \bibfield  {author} {\bibinfo {author} {\bibfnamefont {P.~S.}\ \bibnamefont
  {Skardal}}\ and\ \bibinfo {author} {\bibfnamefont {A.}~\bibnamefont
  {Arenas}},\ }\href@noop {} {\bibfield  {journal} {\bibinfo  {journal}
  {Communications Physics}\ }\textbf {\bibinfo {volume} {3}},\ \bibinfo {pages}
  {218} (\bibinfo {year} {2020})}\BibitemShut {NoStop}%
\bibitem [{\citenamefont {Lee}\ \emph {et~al.}(2024)\citenamefont {Lee},
  \citenamefont {Braun}, \citenamefont {Bönisch}, \citenamefont {Schröder},
  \citenamefont {Thümler},\ and\ \citenamefont {Timme}}]{Seungjae2024}%
  \BibitemOpen
  \bibfield  {author} {\bibinfo {author} {\bibfnamefont {S.}~\bibnamefont
  {Lee}}, \bibinfo {author} {\bibfnamefont {L.}~\bibnamefont {Braun}}, \bibinfo
  {author} {\bibfnamefont {F.}~\bibnamefont {Bönisch}}, \bibinfo {author}
  {\bibfnamefont {M.}~\bibnamefont {Schröder}}, \bibinfo {author}
  {\bibfnamefont {M.}~\bibnamefont {Thümler}}, \ and\ \bibinfo {author}
  {\bibfnamefont {M.}~\bibnamefont {Timme}},\ }\href@noop {} {\bibfield
  {journal} {\bibinfo  {journal} {Chaos: An Interdisciplinary Journal of
  Nonlinear Science}\ }\textbf {\bibinfo {volume} {34}},\ \bibinfo {pages}
  {053141} (\bibinfo {year} {2024})}\BibitemShut {NoStop}%
\bibitem [{\citenamefont {Daido}(1996)}]{daido1996onset}%
  \BibitemOpen
  \bibfield  {author} {\bibinfo {author} {\bibfnamefont {H.}~\bibnamefont
  {Daido}},\ }\href@noop {} {\bibfield  {journal} {\bibinfo  {journal} {Physica
  D: Nonlinear Phenomena}\ }\textbf {\bibinfo {volume} {91}},\ \bibinfo {pages}
  {24} (\bibinfo {year} {1996})}\BibitemShut {NoStop}%
\bibitem [{\citenamefont {Ott}\ and\ \citenamefont {Antonsen}(2008)}]{Ott2008}%
  \BibitemOpen
  \bibfield  {author} {\bibinfo {author} {\bibfnamefont {E.}~\bibnamefont
  {Ott}}\ and\ \bibinfo {author} {\bibfnamefont {T.~M.}\ \bibnamefont
  {Antonsen}},\ }\href {\doibase 10.1063/1.2930766} {\bibfield  {journal}
  {\bibinfo  {journal} {Chaos}\ }\textbf {\bibinfo {volume} {18}},\ \bibinfo
  {pages} {1} (\bibinfo {year} {2008})},\ \Eprint
  {http://arxiv.org/abs/arXiv:0806.0004v1} {arXiv:arXiv:0806.0004v1}
  \BibitemShut {NoStop}%
\bibitem [{\citenamefont {Buzanello}\ \emph {et~al.}(2022)\citenamefont
  {Buzanello}, \citenamefont {Barioni},\ and\ \citenamefont
  {de~Aguiar}}]{buzanello2022matrix}%
  \BibitemOpen
  \bibfield  {author} {\bibinfo {author} {\bibfnamefont {G.~L.}\ \bibnamefont
  {Buzanello}}, \bibinfo {author} {\bibfnamefont {A.~E.~D.}\ \bibnamefont
  {Barioni}}, \ and\ \bibinfo {author} {\bibfnamefont {M.~A.}\ \bibnamefont
  {de~Aguiar}},\ }\href@noop {} {\bibfield  {journal} {\bibinfo  {journal}
  {Chaos: An Interdisciplinary Journal of Nonlinear Science}\ }\textbf
  {\bibinfo {volume} {32}},\ \bibinfo {pages} {093130} (\bibinfo {year}
  {2022})}\BibitemShut {NoStop}%
\bibitem [{\citenamefont {Lohe}(2009)}]{lohe2009non}%
  \BibitemOpen
  \bibfield  {author} {\bibinfo {author} {\bibfnamefont {M.}~\bibnamefont
  {Lohe}},\ }\href@noop {} {\bibfield  {journal} {\bibinfo  {journal} {Journal
  of Physics A: Mathematical and Theoretical}\ }\textbf {\bibinfo {volume}
  {42}},\ \bibinfo {pages} {395101} (\bibinfo {year} {2009})}\BibitemShut
  {NoStop}%
\bibitem [{\citenamefont {Ha}\ and\ \citenamefont
  {Kim}(2019)}]{ha2019emergent}%
  \BibitemOpen
  \bibfield  {author} {\bibinfo {author} {\bibfnamefont {S.-Y.}\ \bibnamefont
  {Ha}}\ and\ \bibinfo {author} {\bibfnamefont {D.}~\bibnamefont {Kim}},\
  }\href@noop {} {\bibfield  {journal} {\bibinfo  {journal} {Journal of
  Statistical Physics}\ }\textbf {\bibinfo {volume} {175}},\ \bibinfo {pages}
  {904} (\bibinfo {year} {2019})}\BibitemShut {NoStop}%
\bibitem [{\citenamefont {Bronski}\ \emph {et~al.}(2020)\citenamefont
  {Bronski}, \citenamefont {Carty},\ and\ \citenamefont
  {Simpson}}]{bronski2020matrix}%
  \BibitemOpen
  \bibfield  {author} {\bibinfo {author} {\bibfnamefont {J.~C.}\ \bibnamefont
  {Bronski}}, \bibinfo {author} {\bibfnamefont {T.~E.}\ \bibnamefont {Carty}},
  \ and\ \bibinfo {author} {\bibfnamefont {S.~E.}\ \bibnamefont {Simpson}},\
  }\href@noop {} {\bibfield  {journal} {\bibinfo  {journal} {Journal of
  Statistical Physics}\ }\textbf {\bibinfo {volume} {178}},\ \bibinfo {pages}
  {595} (\bibinfo {year} {2020})}\BibitemShut {NoStop}%
\bibitem [{\citenamefont {Mehta}(2004)}]{mehta2004random}%
  \BibitemOpen
  \bibfield  {author} {\bibinfo {author} {\bibfnamefont {M.~L.}\ \bibnamefont
  {Mehta}},\ }\href@noop {} {\emph {\bibinfo {title} {Random matrices}}}\
  (\bibinfo  {publisher} {Elsevier},\ \bibinfo {year} {2004})\BibitemShut
  {NoStop}%
\bibitem [{\citenamefont {Meckes}(2019)}]{meckes2019random}%
  \BibitemOpen
  \bibfield  {author} {\bibinfo {author} {\bibfnamefont {E.~S.}\ \bibnamefont
  {Meckes}},\ }\href@noop {} {\emph {\bibinfo {title} {The random matrix theory
  of the classical compact groups}}},\ Vol.\ \bibinfo {volume} {218}\ (\bibinfo
   {publisher} {Cambridge University Press},\ \bibinfo {year}
  {2019})\BibitemShut {NoStop}%
\bibitem [{\citenamefont {Hua}(1963)}]{hua1963harmonic}%
  \BibitemOpen
  \bibfield  {author} {\bibinfo {author} {\bibfnamefont {L.~K.}\ \bibnamefont
  {Hua}},\ }\href@noop {} {\emph {\bibinfo {title} {Harmonic analysis of
  functions of several complex variables in the classical domains}}},\ \bibinfo
  {number} {6}\ (\bibinfo  {publisher} {American Mathematical Soc.},\ \bibinfo
  {year} {1963})\BibitemShut {NoStop}%
\bibitem [{\citenamefont {B{\'e}ri}(2009)}]{beri2009random}%
  \BibitemOpen
  \bibfield  {author} {\bibinfo {author} {\bibfnamefont {B.}~\bibnamefont
  {B{\'e}ri}},\ }\href@noop {} {\bibfield  {journal} {\bibinfo  {journal}
  {Physical Review B}\ }\textbf {\bibinfo {volume} {79}},\ \bibinfo {pages}
  {214506} (\bibinfo {year} {2009})}\BibitemShut {NoStop}%
\bibitem [{\citenamefont {Mello}\ \emph {et~al.}(1985)\citenamefont {Mello},
  \citenamefont {Pereyra},\ and\ \citenamefont
  {Seligman}}]{mello1985information}%
  \BibitemOpen
  \bibfield  {author} {\bibinfo {author} {\bibfnamefont {P.~A.}\ \bibnamefont
  {Mello}}, \bibinfo {author} {\bibfnamefont {P.}~\bibnamefont {Pereyra}}, \
  and\ \bibinfo {author} {\bibfnamefont {T.~H.}\ \bibnamefont {Seligman}},\
  }\href@noop {} {\bibfield  {journal} {\bibinfo  {journal} {Annals of
  Physics}\ }\textbf {\bibinfo {volume} {161}},\ \bibinfo {pages} {254}
  (\bibinfo {year} {1985})}\BibitemShut {NoStop}%
\bibitem [{\citenamefont {Doron}\ and\ \citenamefont
  {Smilansky}(1992)}]{doron1992some}%
  \BibitemOpen
  \bibfield  {author} {\bibinfo {author} {\bibfnamefont {E.}~\bibnamefont
  {Doron}}\ and\ \bibinfo {author} {\bibfnamefont {U.}~\bibnamefont
  {Smilansky}},\ }\href@noop {} {\bibfield  {journal} {\bibinfo  {journal}
  {Nuclear Physics, A;(Netherlands)}\ }\textbf {\bibinfo {volume} {545}}
  (\bibinfo {year} {1992})}\BibitemShut {NoStop}%
\bibitem [{\citenamefont {Baranger}\ and\ \citenamefont
  {Mello}(1996)}]{baranger1996short}%
  \BibitemOpen
  \bibfield  {author} {\bibinfo {author} {\bibfnamefont {H.}~\bibnamefont
  {Baranger}}\ and\ \bibinfo {author} {\bibfnamefont {P.}~\bibnamefont
  {Mello}},\ }\href@noop {} {\bibfield  {journal} {\bibinfo  {journal}
  {Europhysics Letters}\ }\textbf {\bibinfo {volume} {33}},\ \bibinfo {pages}
  {465} (\bibinfo {year} {1996})}\BibitemShut {NoStop}%
\bibitem [{\citenamefont {Brouwer}(1995)}]{brouwer1995generalized}%
  \BibitemOpen
  \bibfield  {author} {\bibinfo {author} {\bibfnamefont {P.~W.}\ \bibnamefont
  {Brouwer}},\ }\href@noop {} {\bibfield  {journal} {\bibinfo  {journal}
  {Physical Review B}\ }\textbf {\bibinfo {volume} {51}},\ \bibinfo {pages}
  {16878} (\bibinfo {year} {1995})}\BibitemShut {NoStop}%
\bibitem [{\citenamefont {Jarosz}\ \emph {et~al.}(2015)\citenamefont {Jarosz},
  \citenamefont {Vidal},\ and\ \citenamefont {Kanzieper}}]{jarosz2015random}%
  \BibitemOpen
  \bibfield  {author} {\bibinfo {author} {\bibfnamefont {A.}~\bibnamefont
  {Jarosz}}, \bibinfo {author} {\bibfnamefont {P.}~\bibnamefont {Vidal}}, \
  and\ \bibinfo {author} {\bibfnamefont {E.}~\bibnamefont {Kanzieper}},\
  }\href@noop {} {\bibfield  {journal} {\bibinfo  {journal} {Physical Review
  B}\ }\textbf {\bibinfo {volume} {91}},\ \bibinfo {pages} {180203} (\bibinfo
  {year} {2015})}\BibitemShut {NoStop}%
\bibitem [{\citenamefont {Pikovsky}\ and\ \citenamefont
  {Rosenblum}(2008)}]{pikovsky2008partially}%
  \BibitemOpen
  \bibfield  {author} {\bibinfo {author} {\bibfnamefont {A.}~\bibnamefont
  {Pikovsky}}\ and\ \bibinfo {author} {\bibfnamefont {M.}~\bibnamefont
  {Rosenblum}},\ }\href@noop {} {\bibfield  {journal} {\bibinfo  {journal}
  {Physical review letters}\ }\textbf {\bibinfo {volume} {101}},\ \bibinfo
  {pages} {264103} (\bibinfo {year} {2008})}\BibitemShut {NoStop}%
\bibitem [{\citenamefont {Marvel}\ \emph {et~al.}(2009)\citenamefont {Marvel},
  \citenamefont {Mirollo},\ and\ \citenamefont
  {Strogatz}}]{marvel2009identical}%
  \BibitemOpen
  \bibfield  {author} {\bibinfo {author} {\bibfnamefont {S.~A.}\ \bibnamefont
  {Marvel}}, \bibinfo {author} {\bibfnamefont {R.~E.}\ \bibnamefont {Mirollo}},
  \ and\ \bibinfo {author} {\bibfnamefont {S.~H.}\ \bibnamefont {Strogatz}},\
  }\href@noop {} {\bibfield  {journal} {\bibinfo  {journal} {Chaos: An
  Interdisciplinary Journal of Nonlinear Science}\ }\textbf {\bibinfo {volume}
  {19}} (\bibinfo {year} {2009})}\BibitemShut {NoStop}%
\bibitem [{\citenamefont {Chen}\ \emph {et~al.}(2017)\citenamefont {Chen},
  \citenamefont {Engelbrecht},\ and\ \citenamefont
  {Mirollo}}]{chen2017hyperbolic}%
  \BibitemOpen
  \bibfield  {author} {\bibinfo {author} {\bibfnamefont {B.}~\bibnamefont
  {Chen}}, \bibinfo {author} {\bibfnamefont {J.~R.}\ \bibnamefont
  {Engelbrecht}}, \ and\ \bibinfo {author} {\bibfnamefont {R.}~\bibnamefont
  {Mirollo}},\ }\href@noop {} {\bibfield  {journal} {\bibinfo  {journal}
  {Journal of Physics A: Mathematical and Theoretical}\ }\textbf {\bibinfo
  {volume} {50}},\ \bibinfo {pages} {355101} (\bibinfo {year}
  {2017})}\BibitemShut {NoStop}%
\end{thebibliography}
%

\end{document}